\documentclass[acmsmall,screen,authorversion]{acmart}

\usepackage{booktabs} 

\usepackage[ruled]{algorithm2e} 

\SetAlFnt{\small}
\SetAlCapFnt{\small}
\SetAlCapNameFnt{\small}
\SetAlCapHSkip{0pt}
\IncMargin{-\parindent}

\usepackage{subfigure}
\usepackage{multirow}

\acmJournal{TIIS}

\setcopyright{acmcopyright}

\begin{document}
\title[A Classification Model for Sensing Human Trust in Machines]{A Classification Model for Sensing Human Trust in Machines Using EEG and GSR}  

\author{Kumar Akash}
\author{Wan-Lin Hu}
\author{Neera Jain}
\author{Tahira Reid}
\affiliation{%
  \institution{Purdue University}
  \streetaddress{610 Purdue Mall}
  \city{West Lafayette}
  \state{IN}
  \postcode{47907}
  \country{USA}}

\begin{abstract}
Today, intelligent machines \emph{interact and collaborate} with humans in a way that demands a greater level of trust between human and machine. A first step towards building intelligent machines that are capable of building and maintaining trust with humans is the design of a sensor that will enable machines to estimate human trust level in real-time. In this paper, two approaches for developing classifier-based empirical trust sensor models are presented that specifically use electroencephalography (EEG) and galvanic skin response (GSR) measurements. Human subject data collected from 45 participants is used for feature extraction, feature selection, classifier training, and model validation. The first approach considers a general set of psychophysiological features across all participants as the input variables and trains a classifier-based model for each participant, resulting in a trust sensor model based on the general feature set (i.e., a "general trust sensor model"). The second approach considers a customized feature set for each individual and trains a classifier-based model using that feature set, resulting in improved mean accuracy but at the expense of an increase in training time. This work represents the first use of real-time psychophysiological measurements for the development of a human trust sensor. Implications of the work, in the context of trust management algorithm design for intelligent machines, are also discussed.
\end{abstract}

%
%


%
%


\thanks{This material is based upon work supported by the National Science Foundation under Award No. 1548616. Any opinions, findings, and conclusions or recommendations expressed in this material are those of the author(s) and do not necessarily reflect the views of the National Science Foundation.

Author's addresses: K. Akash, W.-L. Hu, N. Jain {and} T. Reid,
School of Mechanical Engineering, Purdue University, West Lafayette, Indiana 47907.}

\maketitle

\renewcommand{\shortauthors}{K. Akash et al.}

\newcommand{\psycho}{psychophysiological}

\section{Introduction}
Intelligent machines, and more broadly, intelligent systems are becoming increasingly common in the everyday lives of humans. Nonetheless, despite significant advancements in automation, human supervision and intervention are still essential in almost all sectors, ranging from manufacturing and transportation to disaster-management and healthcare~\cite{wang2017trends}. Therefore, we expect that the future will be built around \emph{Human-Agent Collectives}~\cite{Jennings2014} that will require efficient and successful coordination and collaboration between humans and machines.

\begin{figure}
\centerline{\includegraphics[width=0.9\textwidth]{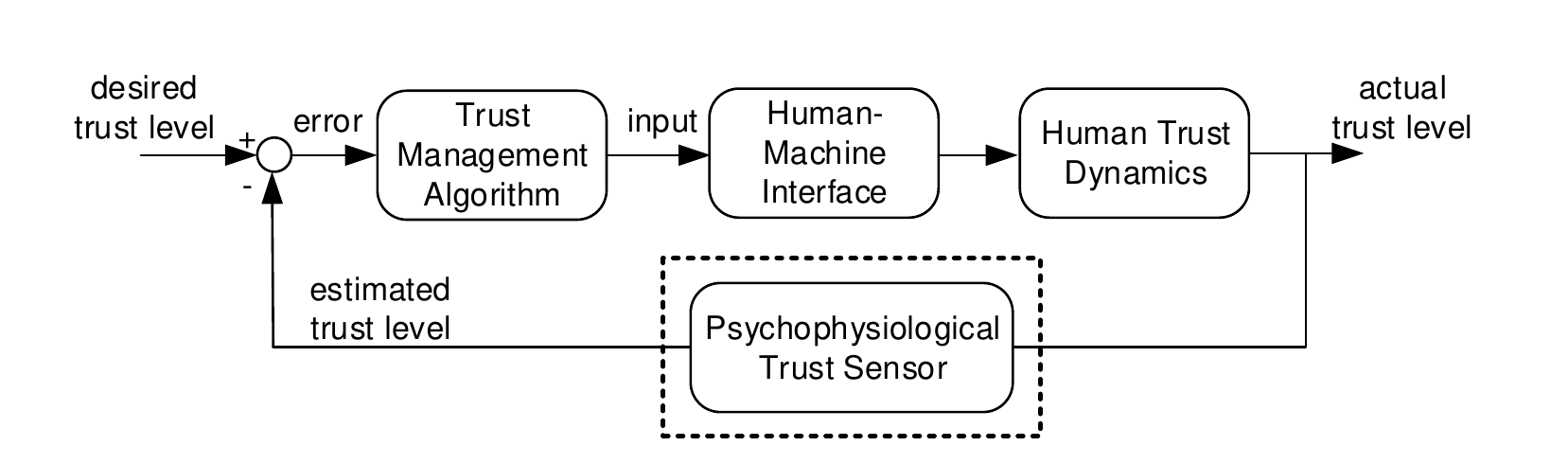}}
\caption{A block diagram of a feedback control system for achieving trust management during human-machine interactions.  The scope of this work includes \psycho~trust sensor modeling.}
\label{fig:block_diagram}
\end{figure}

It is well established that human \emph{trust} is central to successful interactions between humans and machines~\cite{muir_trust_1987,lee2004,sheridan2005}. In the context of autonomous systems, human trust can be classified into three categories: dispositional, situational, and learned~\cite{hoff2015}. Dispositional trust refers to the component of trust that is dependent on demographics such as gender and culture, whereas situational and learned trust depend on a given situation (e.g., task difficulty) and past experience (e.g., machine reliability), respectively. While all of these trust factors influence the way humans make decisions while interacting with intelligent machines, situational and learned trust factors ``can change within the course of a single interaction'' 
~\cite{hoff2015}. Therefore, we are interested in using feedback control principles to design machines that are capable of \emph{responding to changes in human trust level in real-time} to build and manage trust in the human-machine relationship as shown in Figure~\ref{fig:block_diagram}. However, in order to do this, we require a sensor for \emph{estimating human trust level}, again in real-time.

Researchers have attempted to predict human trust using dynamic models that rely on the experience and/or self-reported behavior of humans~\cite{lee1992,Jonker1999}. However, it is not practical to retrieve human self-reported behavior continuously for use in a feedback control algorithm. An alternative is the use of \psycho~signals to estimate trust level~\cite{riedl2012}. While these measurements have been correlated to human trust level~\cite{boudreau2008,Long2012}, they have not been studied in the context of real-time trust sensing.

In this paper we present a human trust sensor model based upon real-time \psycho~measurements, primarily galvanic skin response (GSR) and electroencephalography (EEG). The model is based upon data collected through a human subject study and the use of classification algorithms to estimate human trust level using \psycho~data. The proposed methodology for real-time sensing of human trust level will enable the development of a machine algorithm aimed at improving interactions between humans and machines. 

This paper is organized as follows. In Section~\ref{sec:background} we introduce related work in human-machine interaction, \psycho~measurements, and their applications in trust sensing. We then describe the experimental study and data acquisition in Section~\ref{sec:experiment}. The data pre-processing technique for noise removal is presented in Section~\ref{sec:DataAnalysis} along with EEG and GSR feature extraction. In Section~\ref{sec:FeatureSelection}, we demonstrate a 2-step feature selection process to obtain a concise and optimal feature set. The selected features are then used for training Quadratic Discriminant Analysis classifiers in Section~\ref{sec:modeling}, followed by model validation and finally, concluding statements.
\section{Background and Related Work} \label{sec:background}
There are few \psycho~measurements that have been studied in the context of human trust. We focus here on electroencephalography (EEG) and galvanic skin response (GSR) which are both noninvasive and whose measurements can be collected and processed in real-time. EEG is an electrophysiological measurement technique that captures the cortical activity of the brain~\cite{handy2005}. These brain activities exhibit changes in human thoughts, actions, and emotions. Brain-Computer Interface (BCI) technology utilizes EEG to design interfaces that enable a computer or an electronic device to understand a human's  commands~\cite{Penny2000,Pfurtscheller1993}. The most extensive approach used to identify EEG patterns in BCI design includes feature selection and classification algorithms as they typically provide good accuracy~\cite{Mcfarland2006}. 

Some researchers have studied trust via EEG measurements, but only with event-related potentials (ERPs). ERPs measure brain activity in response to a specific event. An ERP is determined by averaging repeated EEG responses over many trials to eliminate random brain activity~\cite{handy2005}. Boudreau \textit{et al.} found a difference in peak amplitudes of ERP components in human subjects while they participated in a coin toss experiment that stimulated trust and distrust~\cite{boudreau2008}. Long \textit{et al.} further studied ERP waveforms with feedback stimuli based on a modified form of the coin toss experiment~\cite{Long2012}. The decision-making in the ``trust game''~\cite{ma2015} has been used to examine human-human trust level. Although ERPs can show how the brain functionally responds to a stimulus, they are event-triggered. It is difficult to identify triggers during the course of an actual human-machine interaction, thereby rendering ERPs impractical for real-time trust level sensing.

GSR is a classical \psycho~signal that captures arousal based upon the conductivity of the surface of the skin. It is not under conscious control but is instead modulated by the sympathetic nervous system. GSR has also been used in measuring stress, anxiety, and cognitive load ~\cite{Nikula1991,Jacobs1994}. Researchers have examined GSR in correlation with human trust level. Khawaji \textit{et al.} found that average GSR values, and average GSR peak values, are significantly affected by both trust and cognitive load in the text-chat environment~\cite{khawaji_using_2015}. However, the use of GSR for \emph{estimating} trust has not been explored and was noted as an area worth studying~\cite{riedl2012}. With respect to both GSR and EEG, a fundamental gap remains in determining a static model that not only estimates human trust level using these \psycho~signals but that is also suitable for real-time implementation. 
\section{Methods and Procedures} \label{sec:experiment}
\begin{figure}
\centerline{\includegraphics[width=0.9\textwidth]{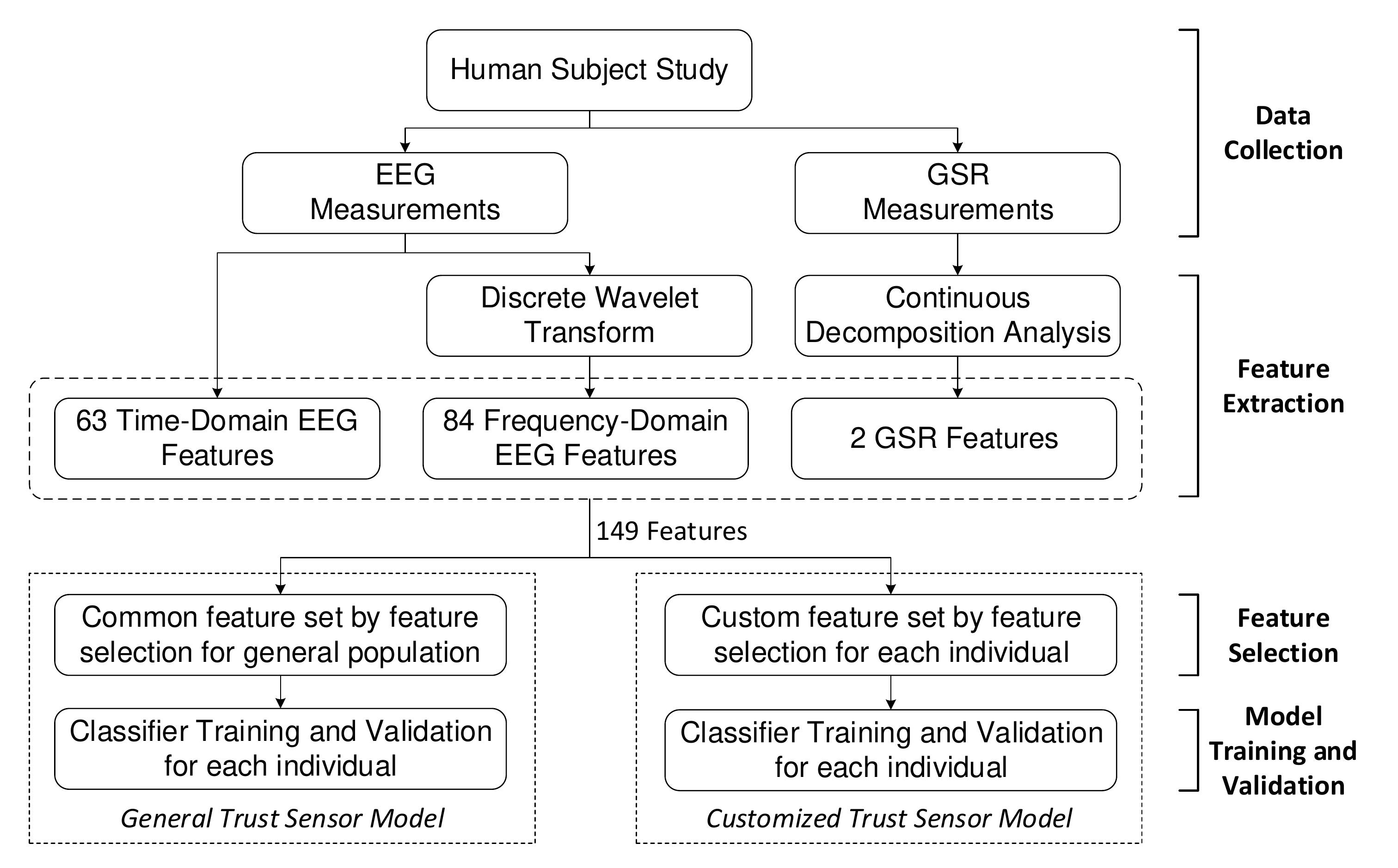}}
\caption{The framework of the proposed study. The key steps include data collection from human subject studies, feature extraction, feature selection, model training, and model validation.}
\label{fig:framework}
\end{figure}

In this section we describe a human subject study that we conducted to identify \psycho~features that are significantly correlated to human trust in intelligent systems, and to build a trust sensor model accordingly. The experiment consisted of a simple HMI context that could elicit human trust dynamics in a simulated autonomous system. Our study used a within-subjects design wherein both behavioral and \psycho~data were collected and analyzed. We then used the data to build an empirical model of human trust through a process involving feature extraction, feature selection, and model training, that is described in Sections~\ref{sec:DataAnalysis}, \ref{sec:FeatureSelection}, and \ref{sec:modeling}, respectively. Figure~\ref{fig:framework} summarizes the modeling framework. 

\subsection{Participants}
Participants were recruited using fliers and email lists.  All participants were compensated at a rate of \$15/hr. The sample included forty-eight adults between 18 and 46 years of age (mean: 25.0 years old, standard deviation: 6.9 years old) from West Lafayette, Indiana (USA). Of the forty-eight adults, sixteen were females and thirty-two were males. All participants were healthy and one was left-handed. The group of participants were diverse with respect to their age, professional field, and cultural background (i.e., nationality). The Institutional Review Board at Purdue University approved the study.

\subsection{EEG and GSR Recording}
\subsubsection*{EEG}
The participant's brain waves were measured using a B-Alert X-10 9-channel EEG device (Advance Brain Monitoring, CA, USA), at a frequency of 256 Hz from 9 scalp sites (Fz, F3, F4, Cz, C3, C4, POz, P3, and P4 based on the 10-20 system). All EEG channels were referenced to the mean of the left and right mastoids. The surface of all sensor sites was cleaned with 70\% isopropyl alcohol. Conductive electrode cream (Kustomer Kinetics, CA, USA) was then applied to each electrode including the reference. The contact impedance between electrodes and skin was kept to a value less than 40~k$\Omega$. The EEG signal was recorded via iMotions (iMotions, Inc., MA, USA) on a Windows 7 platform with Bluetooth connection. 

\subsubsection*{GSR}
The skin conductance was measured from the proximal phalanges of the index and the middle fingers of the non-dominant hand (i.e., on the left hand for 43 out of 44 participants) at a frequency of 52 Hz via the Shimmer3 GSR+ Unit (Shimmer, MA, USA). Locations for attaching Ag/AgCl electrodes (Lafayette Instrument, IN, USA) were prepared with 70\% isopropyl alcohol. The participants were asked to keep their hands steady on the desk to minimize the influence of movement on the measured signals. The environment temperature was controlled at 72-74$^{\circ}$F to minimize the effect of temperature. The GSR signal was also recorded via iMotions so that it would be synchronized with the recorded EEG signals using the common system-timestamps between these two signals. 

\subsection{Experimental Procedure}
\begin{figure}
\centerline{\includegraphics[width=0.6\textwidth]{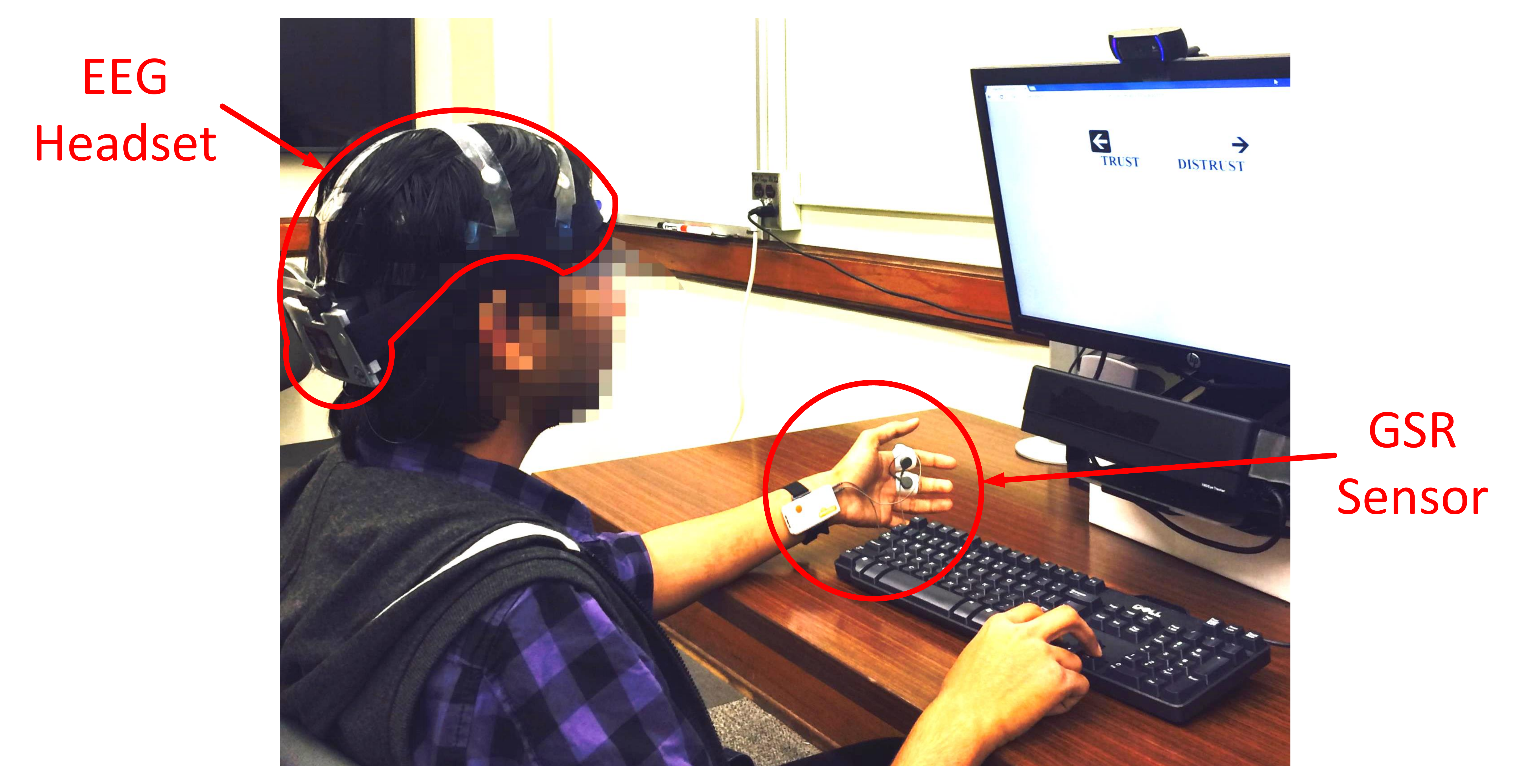}}
\caption{Experimental setup with participant wearing EEG Headset and GSR Sensor.}
\label{fig:exp_setup}
\end{figure}

After the participants read and signed the informed consent, they were equipped with the EEG headset and the GSR sensor as shown in Figure~\ref{fig:exp_setup}. All participants finished a 9-minute EEG baseline task provided by Advanced Brain Monitoring and were then instructed to interact with our custom-designed computer-based simulation. Participants were told that they would be driving a car equipped with an image--based obstacle detection sensor. The sensor would detect obstacles on the road in front of the car, and the participant would need to repeatedly evaluate the algorithm report and choose to either trust or distrust the report based on their experience with the algorithm. Detailed instructions were  delivered on the screen following four practice trials. Participants could have their questions answered while instructions were given and during the practice session.

\begin{figure}
\centerline{\includegraphics[width=.9\textwidth]{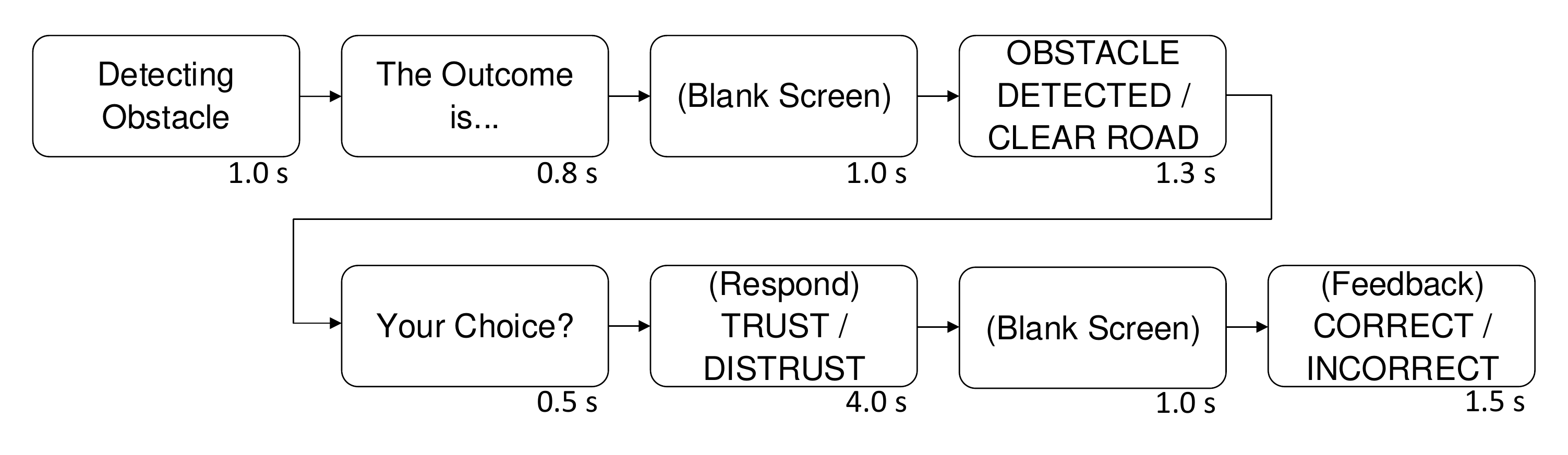}}
\caption{Sequence of events in a single trial. The time length marked on the bottom right corner of each event indicates the time interval for which the information appeared on the computer screen.}
\label{fig:study_design}
\end{figure}

\begin{figure}
\centering
\subfigure[Stimuli\label{subfig:screenshot_stimulus}]{\includegraphics[width=0.3\textwidth]{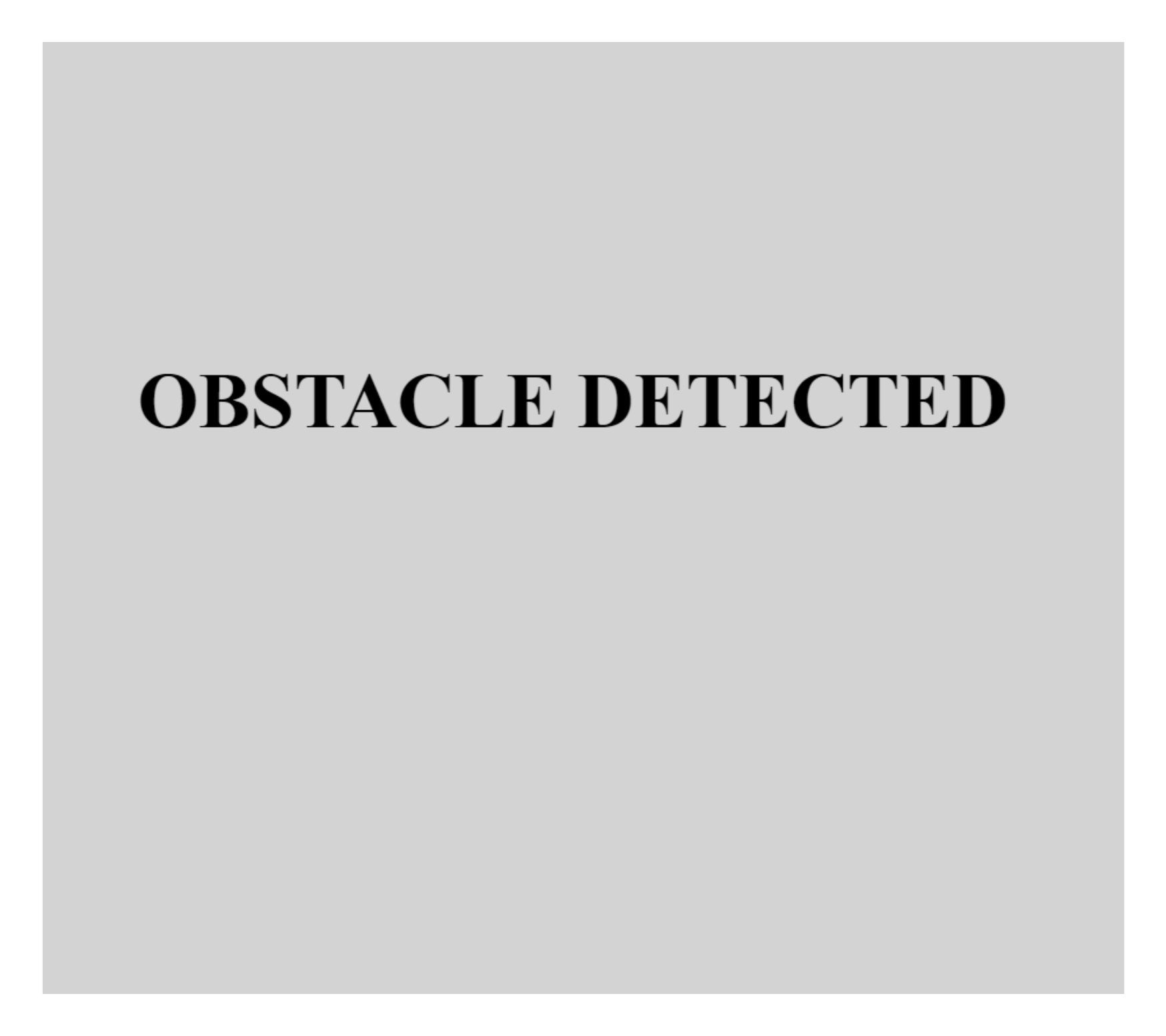}}
\hfill
\subfigure[Response\label{subfig:screenshot_response}]{\includegraphics[width=0.3\textwidth]{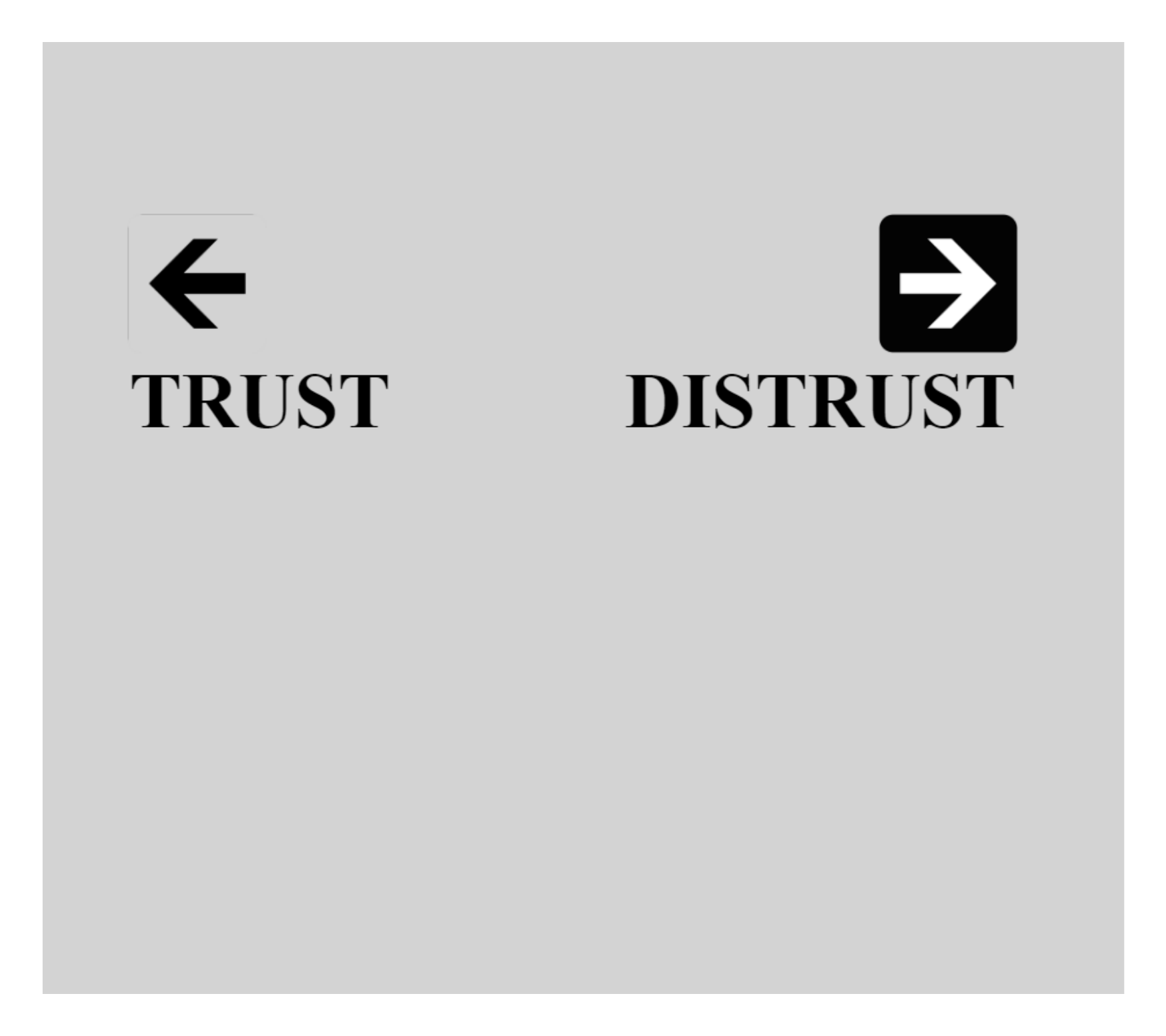}}
\hfill
\subfigure[Feedback\label{subfig:screenshot_feedback}]{\includegraphics[width=0.3\textwidth]{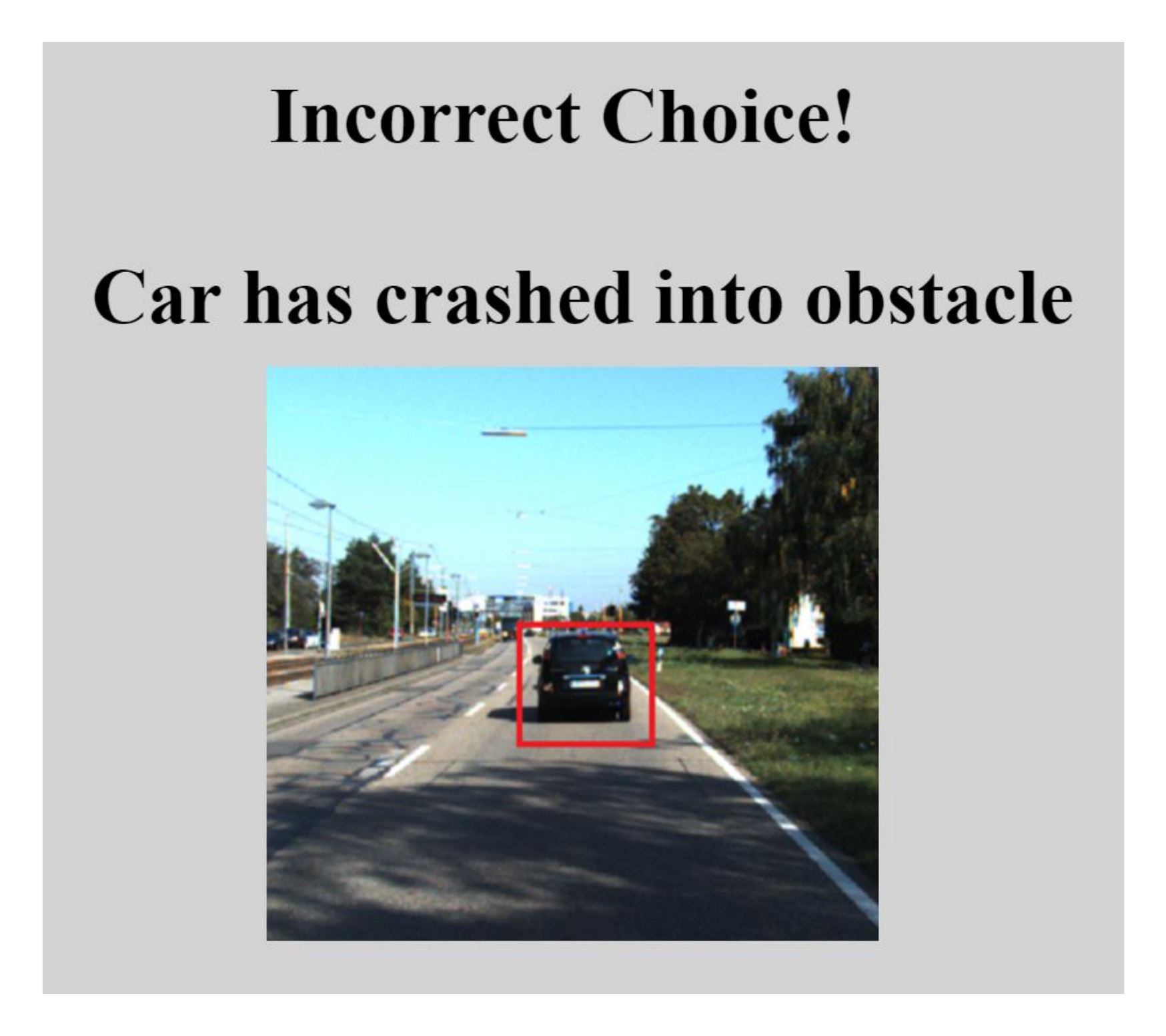}}
\caption{Example screenshots of the interface of the experimental study. These screens correspond to three of the events shown in Figure~\ref{fig:study_design}: obstacle detected/clear road, trust/distrust, and correct/incorrect, respectively.}
\label{fig:screenshot}
\end{figure}

Each trial consisted of: a stimulus (i.e., report on sensor functionality), the participant's response, and feedback to the participants on the correctness of their response. There were two stimuli, `obstacle detected' and `clear road', and both had a 50\% probability of occurrence. Participants had the option to choose `trust' or `distrust' in response to the sensor report after which they received the feedback of `correct' or `incorrect'. Figure~\ref{fig:study_design} shows the sequence of events in a single trial, and Figure~\ref{fig:screenshot} shows example screenshots of the computer interface. 

The independent variable was the participants' experience due to the sensor performance, and the dependent variable was their trust level. The sensor performance was varied to elicit the dynamic response in each participant's trust level. There were two categories of trials: \emph{reliable} and \emph{faulty}. In reliable trials, the sensor accurately identified the road condition with 100\% probability; in faulty trials, there was only a 50\% probability that the sensor correctly identified the road condition with sensor faults presented in a randomized order. We implemented the 50\% accuracy for faulty trials because pilot studies indicated that it would be perceived as a pure random chance by the participants. This should conceivably result in the lowest possible trust level that a human has in the simulated sensor. The participants received  `correct' as feedback when they indicated trust in reliable trials, but there was a 50\% probability that they received `incorrect' as feedback when they indicated trust in faulty trials. 

\begin{figure}
\centerline{\includegraphics[width=0.8\textwidth]{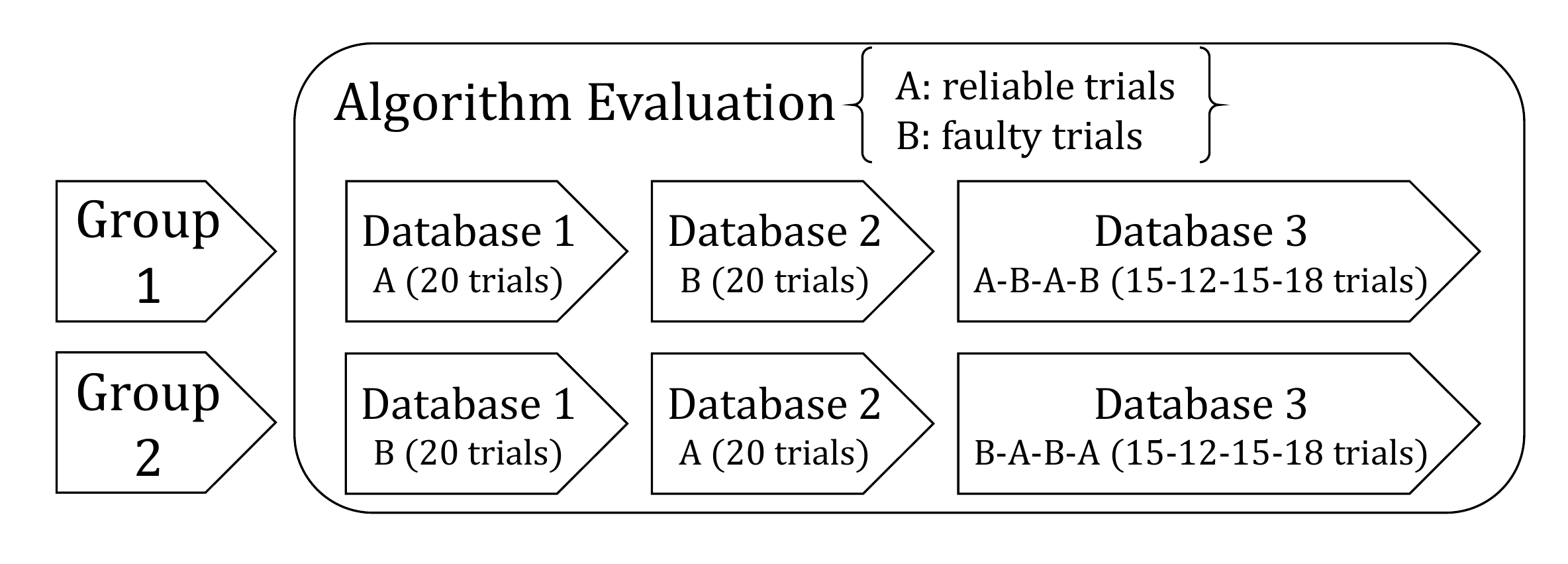}}
\caption{Participants were randomly assigned to one of two groups. The ordering of the three experimental sections (databases), composed of reliable and faulty trials, were counterbalanced across Groups 1 and 2.}
\label{fig:trials}
\end{figure}

Each participant completed 100 trials. The trials were divided into three phases, called `databases' in the study, as shown in Figure~\ref{fig:trials}. Participants were randomly assigned to one of  two groups for counterbalancing any possible ordering effects. Databases 1 and 2 consisted of either reliable (A) or faulty (B) trials (see details in Figure~\ref{fig:trials}). The number of trials in each of these two databases was chosen so that the trust or distrust response of each human subject would approach a steady-state value~\cite{Long2012}. Steady-state ensures that the trust level truly reaches the desired state (i.e., trust for reliable trials and distrust for faulty trials) which is essential for labeling the trials as trust or distrust. On the other hand, the accuracy of the algorithm was switched between reliable and faulty according to a pseudo-random binary sequence (PRBS) in Database 3. This was done in order to excite all possible dynamics of the participant's trust response required for dynamic behavior modeling, which was the subject of related work by the authors~\cite{Akash2017}. Therefore, only the data from databases 1 and 2 (i.e., the first 40 trials) were analyzed.

\begin{figure}
\centering
\subfigure[Group 1; 295 participants\label{fig:mturk1}]{\includegraphics[width=0.7\textwidth]{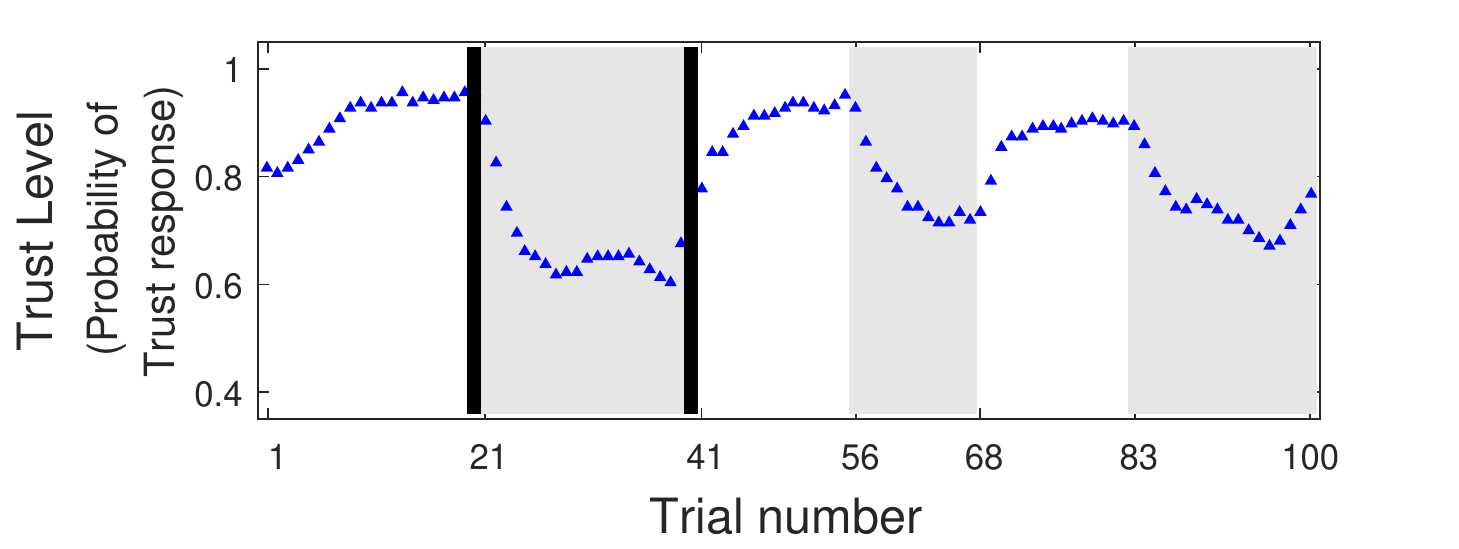}}
\subfigure[Group 2; 228 participants\label{fig:mturk2}]{\includegraphics[width=0.7\textwidth]{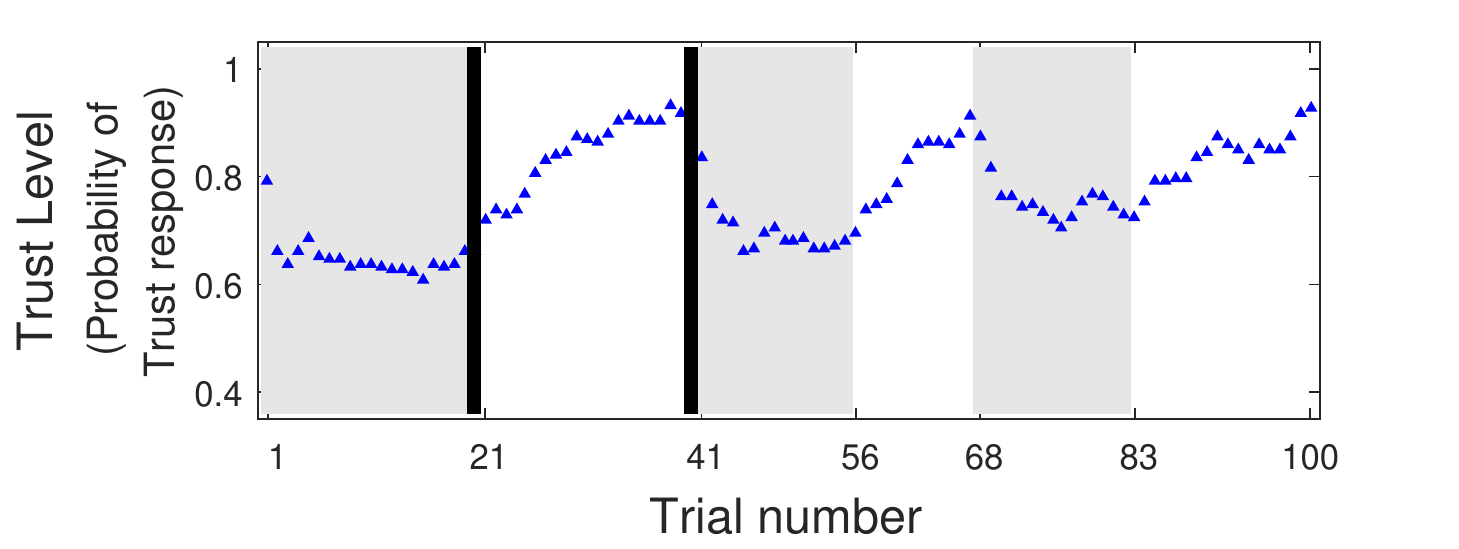}}
\caption{The averaged response from online participants collected via Amazon Mechanical Turk. Faulty trials are highlighted in gray. Participants showed a high trust level in reliable trials and a low trust level in faulty trials regardless of the group they were in.}
\label{fig:mturk}
\end{figure}

We collected \psycho~measurements in order to identify any latent indicators of trust and distrust. In general, latent emotions are those which cannot be easily articulated. Latent distrust may inhibit the interactions between human and intelligent systems despite reported trust behaviors.  We hypothesized that the trust level would be high in reliable trials and be low in faulty trials, and we validated this hypothesis using responses collected from 581 online participants (58 were outliers) via Amazon Mechanical Turk~\cite{mturk_amazon_2005}. The experiment elicited expected trust responses based on the aggregated data as shown in Figure~\ref{fig:mturk}~\cite{Akash2017}. Therefore, data from reliable trials were labeled as trust, and data from faulty trials were labeled as distrust. The data analysis and feature extraction methodologies will be discussed further in Section~\ref{sec:DataAnalysis}. 

\section{Data Analysis} \label{sec:DataAnalysis}
In this section we discuss the methods used to pre-process the data (collected during the human subject studies) so as to reduce noise and remove contaminated data. We then describe the process of feature extraction applied to the processed data.

\subsection{Pre-processing}
We used the automatic decontaminated signals provided by the B-Alert EEG system for artifact removal. This decontamination process minimizes the effects of electromyography, electrooculography, spikes, saturation, and excursions. Before further processing the data, we manually examined the spectral distribution of EEG data for each participant. We removed the participants having anomalous EEG spectra, possibly due to bad channels or dislocation of EEG electrodes during the study. This process resulted in 45 participants to analyze. Finally, EEG measurements from channel F3 and F4 were excluded from the data analysis due to contamination with eye movement and blinking~\cite{berka2007}. For GSR measurements, we used adaptive Gaussian smoothing with a window of size 8 to reduce noise~\cite{blinchikoff1976}.

\subsection{Feature Extraction} \label{subsec:FeatureExtraction}
In order to estimate trust in real-time, we require the ability to continuously extract and evaluate key \psycho~measurements. This could be achieved by continuously considering short segments of signals for calculations. Levy suggests using short epoch lengths for identifying rapid changes in EEG patterns~\cite{Levy1987}. Therefore, we divided the entire duration of the study into multiple 1-second epochs (periods) with 50\% overlap between each consecutive epoch. Assuming that the decisive cognitive activity occurs when the participant sees the stimuli, we only considered the epochs lying completely between each successive stimulus (obstacle detected/clear road) and response (trust/distrust). Consequently, approximately 129 epochs were considered for each participant. We labeled each of these epochs as one of two classes, namely \emph{Distrust} or \emph{Trust}, based on whether the epoch belonged to faulty or reliable trials, respectively. The number of epochs varied depending on the response time of the human subject for each trial.

\subsubsection*{EEG} 
Existing studies have shown the importance of both time-domain features and frequency-domain features for successfully classifying cognitive tasks~\cite{Lotte2007}. To utilize the benefits of both, we extracted an exhaustive set of time- and frequency-domain features from EEG. 

We extracted six time-domain features from all seven channels (Fz, C3, Cz, C4, P3, POz, and P4) for each epoch of length $N$. For this study in which EEG signals were sampled at 256 Hz, each 1-second epoch had a length of $N=256$. Letting $k\in\left(1,n\right)$, where $n$ is the total number of epochs and $x_k$ represents the $k^{th}$ epoch of channel $ch_x$. These features were defined as:
\begin{enumerate}
\item mean $\mu_k(ch_x)$, where
\begin{equation}
\label{eq:mean}
\mu_k(ch_x) = \frac{1}{N}\sum\limits_{i=1}^N x_{ki} \text{,}
\end{equation}

\item variance $\sigma^2_k(ch_x)$, where
\begin{equation}
\label{eq:var}
\sigma^2_k(ch_x) = \frac{1}{N-1}\sum\limits_{i=1}^N |x_{ki} - \mu_k|^2 \text{,}
\end{equation}

\item peak-to-peak value $pp_k(ch_x)$, where
\begin{equation}
\label{eq:p2p}
pp_k(ch_x) = \max\limits_{1\leq i\leq N}x_{ki} - \min\limits_{1\leq i\leq N}x_{ki} \text{,}
\end{equation}

\item mean frequency $\bar{f}_k(ch_x)$, defined as the estimate of the mean frequency from the power spectrum of $x_k$,
\bigskip
\item root mean square value $rms_k(ch_x)$, where
\begin{equation}
\label{eq:rms}
rms_k(ch_x) = \sqrt{\frac{1}{N}\sum\limits_{i=1}^N |x_{ki}|^2} \text{,}
\end{equation}
\noindent and
\bigskip
\item signal energy $E_k(ch_x)$, where 
\begin{equation}
\label{eq:energy}
E_k(ch_x) = \sum\limits_{i=1}^N |x_{ki}|^2 \enspace .
\end{equation}

\end{enumerate}

Therefore, we extracted 42 (6 features $\times$ 7 channels) time-domain features for each epoch. Moreover, the interaction between the different regions of the brain was also considered by calculating the correlation between pairs of channels for each epoch. The correlation coefficient between two channels (e.g., $ch_x$ and $ch_y$) of the $k^{th}$ epoch $\rho_k(ch_x,ch_y)$ is defined as 

\begin{equation}
\label{eq:corr}
\rho_k(ch_x,ch_y) = \frac{cov(x_k,y_k))}{\sqrt{var(x_k)var(y_k)}},
\end{equation}
where $x_k$ and $y_k$ are the $k^{th}$ epochs of channels $ch_x$ and $ch_y$ respectively. The expressions $cov(.)$ and $var(.)$ are the covariance and variance functions, respectively. Therefore, 21  additional time-domain features were extracted (combinations of 2 out of 7 channels, $C^7_2$). 

Next we extracted features from four frequency bands across all seven channels for each epoch. Classically, EEG brain waves have been categorized into four bands based on frequency, namely, delta (0.5 - 4 Hz), theta (4 - 8 Hz), alpha (8 - 13 Hz), and beta (13 - 30 Hz). However, because of the non-stationary characteristics of EEG signals (i.e., their statistics vary in time), analyzing the variations in frequency components of EEG signal with time (i.e., time-frequency analysis) is more informative than analyzing the frequency content of the entire signal at a time. The Discrete Wavelet Transform (DWT) is an extensively used tool for time-frequency analysis of physiological signals, including EEG~\cite{Amin2015}. Therefore, we used DWT  decomposition to extract the frequency-domain features from the EEG signals. 

DWT uses scale-varying basis functions to achieve good time resolution of high frequencies and good frequency resolution for low frequencies. The DWT decomposition consists of successive high pass and low pass filtering of the signal with downsampling by a factor of 2 in each successive level~\cite{sundararajan2016}. The high pass filter uses a discrete mother wavelet function, and the low pass filter uses its mirror version. We used the mother wavelet function of the Daubechies wavelet (db5) for frequency decomposition of the EEG signal. The first low pass and high pass filter outputs are called approximation A1 and detailed coefficients D1, respectively. A1 is further decomposed, and the steps are repeated to achieve the desired level of decomposition. Since the highest frequency in our signal was 128 Hz (sampling frequency $f_s =$ 256 Hz), each channels' signal was decomposed to the fifth level to achieve the decomposition corresponding to the classical bands as shown in Table~\ref{tab:wavedecompose}.

\begin{table}
\caption{Wavelet decompositions and their frequency range\label{tab:wavedecompose}}{
\begin{tabular}{lccc} 
\hline
Level 	&	Wavelet coefficient 	& 	Frequency range & 	Classical band \\
 \hline
 3		&	D3 						&  16 - 32 Hz 	& 	Beta	\\
 4		&	D4 						&  8  - 16 Hz 	& 	Alpha	\\
 5		&	D5 						&  4  - 8  Hz 	& 	Theta	\\
 5		&	A5 						&  0  - 4  Hz 	& 	Delta	\\
 \hline
\end{tabular}}
\end{table}

Three features, namely mean (Equation~\ref{eq:mean}), variance (Equation~\ref{eq:var}), and energy (Equation~\ref{eq:energy}) were calculated from each of the four decomposed band decomposition coefficients shown in Table~\ref{tab:wavedecompose} for each channel's epoch. Therefore, 84 frequency-domain features were extracted (3 features $\times$ 4 bands $\times$ 7 channels).

\subsubsection*{GSR}
GSR is a superposition of the tonic (slow-changing) and the phasic (fast-changing) components of the skin conductance response~\cite{Benedek2010}. We used Continuous Decomposition Analysis from Ledalab to separate the tonic and phasic components of the signal~\cite{Benedek2010}. Since the time-scale of the study and the decision making tasks are, in general, much faster as compared to the tonic component, we only used the phasic component of the GSR. We calculated the \emph{Maximum Phasic Component} and the \emph{Net Phasic Component} for each epoch, thus extracting 2 features from GSR.
\section{Feature Selection} \label{sec:FeatureSelection}
Following the feature extraction described in Section~\ref{sec:DataAnalysis}, we next describe the process of feature selection. The selected features were considered to be potential input variables for the trust sensor model, of which the output would be the \emph{probability of trust response}. We define the probability of trust response as the probability of the human trusting the intelligent system at the next time instant.  In this section we discuss feature selection algorithms used for selecting optimal feature sets for two variations of our trust sensor model, followed by a discussion of the significance of the features in each of the final feature sets. 

\subsection{Feature Selection Algorithms} \label{subsec:feature_algo}
The complete feature set consisted of 149 features (42 + 21 + 84 + 2) that were extracted for each epoch for every participant. These features were considered potential variables for predicting the \emph{Trust} or \emph{Distrust} classes. Out of this large feature set, it was necessary to downselect a smaller subset of features as predictors to avoid `the curse of dimensionality' (also called Hughes phenomenon), which occurs for high-dimensional feature spaces with a limited number of samples. Not doing feature selection leads to a reduction in the predictive power of learning algorithms~\cite{Lotte2007}. Therefore, feature selection was achieved by removing irrelevant and redundant features from the feature set according to feature selection algorithms.

\begin{figure}
\centerline{\includegraphics[width=1\textwidth]{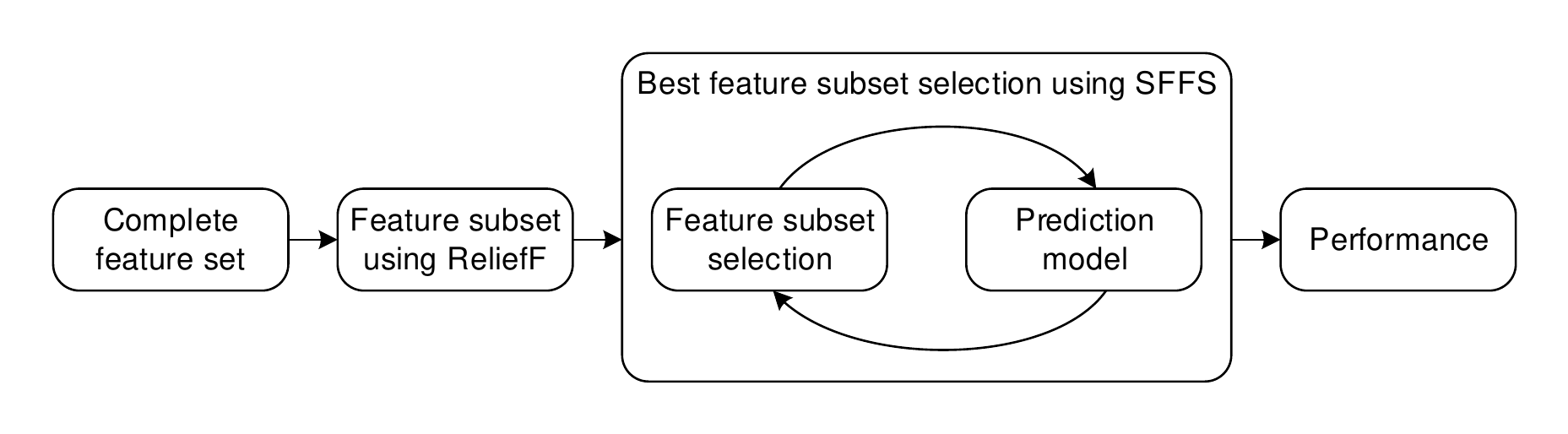}}
\caption{A schematic depicting the feature selection approach used for reducing the dimension of the feature set. The ReliefF (filter method) was used for an initial shortlisting of the feature subset followed by SFFS (wrapper method) for the final feature subset selection.}
\label{fig:feture_selection}
\end{figure}

Feature selection algorithms are categorized into two groups: filter methods and wrapper methods. Filter methods depend on general data characteristics such as inter-class distance, results of significance tests, and mutual information, to select the feature subsets without involving any selected prediction model. Since filter methods do not involve any assumptions of a prediction model, they are useful in estimating the relationships between the features. Wrapper methods use the performance (e.g., accuracy) of a selected prediction model to evaluate possible feature subsets. When the performance of a particular type of model is of importance, wrapper methods result in a better fit for a selected model type; however, they are typically much slower than filter methods \cite{Kohavi1997}. We used a combination of filter and wrapper methods for feature selection to manage the trade-off between training speed and model performance. We used a filter method called \emph{ReliefF} for initially shortlisting features followed by a wrapper method called \emph{Sequential Forward Floating Selection (SFFS)} for the final feature selection as shown in Figure~\ref{fig:feture_selection}. 

\subsubsection{ReliefF}
The basic idea of ReliefF is to estimate the quality of the features based on their ability to distinguish between samples that are near each other. Kononenko \textit{et al.} proposed a number of improvements to existing work by Kira and Rendell and developed ReliefF \cite{Kononenko1997,Kira1992}. For a data set with $n$ samples, the algorithm iterates $n$ times for each feature. For our study, there were approximately 129 samples corresponding to each epoch as mentioned in Section~\ref{subsec:FeatureExtraction}. At each iteration for a two-class problem, the algorithm selects one of the samples and finds $k$ nearest hits (same-class sample) and $k$ nearest misses (different-class sample), where $k$ is a parameter to be selected. Kononenko \textit{et al.} suggested that $k$ could be safely set to 10 for most purposes. We used $k=$10 and calculated the ReliefF weights for all extracted features of each individual participant. The weight of any given feature is penalized for far-off near-hits and improved for far-off near-misses. Far-off near misses implies well-separated features, and far-off near-hits implies intermixed classes.

\subsubsection{Sequential Forward Floating Selection (SFFS)}
The SFFS is an enhancement of the Sequential Feature Selection algorithm for addressing the `nesting effect'~\cite{Pudil1994}. The nesting effect means that a selected feature cannot be discarded when the forward method is implemented and the discarded feature cannot be re-selected when the backward method is implemented. In order to avoid this effect, SFFS builds the feature set with the best predictive power by continuously adding a dynamically changing number of features at each step to the existing subset of features. This operation occurs iteratively until no further increase in performance is observed. In this study we defined the performance as the misclassification rate of the Quadratic Discriminant Analysis (QDA) classifier. We have examined that a QDA classifier achieved the highest accuracy for another data set based on the same experimental setup~\cite{Hu2016}, and its output posterior probability is also suitable for interpreting trust. Therefore, we used the QDA classifier and calculated the misclassification rate using 5-fold cross validation~\cite{hastie2009}. This validation technique randomly divides the data into five sets and predicts each set using a model trained for the remaining four sets.

\subsection{Feature selection for the Trust Sensor Model} \label{subsec:model_features}
The differences between humans could introduce differences in their trust behavior. This leads to two approaches for selecting features for sensing trust level: 1) to select a common set of features for a general population, which results in a \emph{general trust sensor model}; and 2) to select a different set of features for each individual, which results in \emph{customized trust sensor model} for each individual.

\subsubsection{Feature Selection for the General Trust Sensor Model} \label{subsubsec:common-feature}
A general trust sensor model is desirable so that it can be used to reflect trust behavior in a general adult population. This model correlates significant \psycho~features with human trust in intelligent systems based on data obtained from a broad range of adult human subjects. Since a general trust sensor model requires a common list of features for all participants, we randomly divided the participants into two groups: the training-sample participants (33 out of 45 participants), which were used to identify the common list of features, and the validation-sample participants (12 out of 45 participants), which were used to validate the selected list of features. We calculated the median of the ReliefF weights across the training-sample participants for all features. The median was used instead of mean to avoid outliers~\cite{Leys2013}. Finally, we shortlisted features with the top 60 median weights and used SFFS for selecting the final set of features. For each training-sample participant's data, a separate classifier was trained and the average value of the misclassification rate for all training-sample participants was used as the predictive power for feature subsets for SFFS.  We obtained a feature set with 12 features consisting of both time- and frequency-domain features of EEG along with net phasic components of GSR. Table~\ref{tab:feature-selection} shows the final list of selected features for the general trust sensor model using training-sample participants.

\begin{table}
\caption{Features to be used as input variables for the general trust sensor model\label{tab:feature-selection}}{
\begin{tabular}{clll} 
 \hline
 	& Feature 					& Measurement 	& Domain  \\ 
 \hline
 1 	& Mean Frequency - Fz 		& EEG 			& Time \\ 
 2 	& Mean Frequency - C3		& EEG 			& Time \\  
 3 	& Mean Frequency - C4 		& EEG 			& Time \\
 4 	& Peak-to-peak - C3			& EEG 			& Time \\
 5 	& Energy of Theta Band - P3	& EEG 			& Frequency \\
 6 	& Variance of Alpha Band - P4	& EEG		& Frequency \\ 
 7 	& Energy of Beta Band - C4	& EEG			& Frequency \\ 
 8 	& Energy of Beta Band - P3	& EEG			& Frequency \\ 
 9 	& Mean of Beta Band - C3	& EEG			& Frequency \\
10 	& Correlation - C3 \& C4	& EEG			& Time \\
11 	& Correlation - Cz \& C4	& EEG			& Time \\
12 	& Net Phasic Component		& GSR			& Time \\
 \hline
\end{tabular}}
\end{table}

\subsubsection{Feature Selection for the Customized Trust Sensor Model} \label{subsubsec:custom-feature}
We followed a similar approach to that used for feature selection in Section~\ref{subsubsec:common-feature}, but the list of features was selected individually for each of the 45 participants. We used ReliefF weights and shortlisted a separate set of features for each participant consisting of the top 60 weights. Then, for each participant, SFFS was used with the misclassification rate as determined by the quadratic discriminant classifier to select a final set of features from the shortlisted feature set. We obtained a relatively smaller feature set for each individual participant, with an average of 4.33 features in each participant's feature set, as compared to 12 features when all of the participants' data was aggregated into a single data set. Table~\ref{tab:frequent-feature} shows each of the features that are significant for at least four of the participants. We observed that there is great diversity in the significant features for each individual which supports the usage of a customized trust sensor model. However, it is important to note that even within this diversity, more than half of the most common features (e.g., mean frequency at C4) are also significant for the general trust sensor model.

\begin{table}
\caption{The most common features that are significant for at least four participants. Features marked with an asterisk ($^*$) are also significant for the general trust sensor model. \label{tab:frequent-feature}}{
\begin{tabular}{clll} 
 \hline
 	& Feature 						& Measurement 	& Domain  \\ 
 \hline
 1 	& Mean Frequency - POz 		& EEG 			& Time \\
 2 	& Mean Frequency - C4$^*$ 	& EEG 			& Time \\
 3 	& Mean Frequency - P3 		& EEG 			& Time \\
 4 	& Mean Frequency - Fz$^*$	& EEG 			& Time \\ 
 5 	& Mean Frequency - C3$^*$	& EEG 			& Time \\ 
 6 	& Peak-to-peak - C3$^*$		& EEG 			& Time \\
 7 	& Variance of Beta Band - P3	& EEG		& Frequency \\ 
 8 	& Mean of Beta Band - P3	& EEG			& Frequency \\
 9 	& Correlation - Cz \& C4$^*$& EEG			& Time \\
10 	& Net Phasic Component$^*$	& GSR			& Time \\
11 	& Maximum Value of Phasic Activity	& GSR	& Time \\
 \hline
\end{tabular}}
\end{table}

\subsection{Discussion on Significant Features in Trust Sensing} \label{subsec:feature_significance}
Several time-domain EEG features were found to be significant, especially the mean frequency of the EEG power distribution and the correlations between the signals from the central regions of the brain (C3, C4, Cz). Time-domain EEG features have been discovered to be significant in brain activities~\cite{Lotte2007}. Moreover, our observation that activities at sites C3 and C4 play an important role in trust behaviors is supported by existing studies that have suggested that central regions of the brain are related to processes associated with problem complexity~\cite{Jausovec2000}, anxiety in a sustained attention task~\cite{righi_anxiety_2009}, and mental workload~\cite{Dussault2005}.

Among the frequency domain EEG features, the measurements from the left parietal lobe, particularly in a high frequency range (i.e., the beta band), responded most strongly to the discrepancy between reliable and faulty stimuli. This is consistent with the finding that cognitive task demands have a significant interaction with hemisphere in the beta band for parietal areas~\cite{Ray1985}. The beta band is also an important feature that has been shown to be related to emotional states in the literature~\cite{Isotani2001} and may represent the emotional component of human trust. 

Finally, the results also showed that the phasic component of GSR was a significant predictor of trust levels for the general trust sensor model as well as for several customized trust sensor models. This aligns with the existing literature that shows that the GSR features could significantly improve the classification accuracy for mental workload detection~\cite{Chen2012} and could index difficulty levels of decision making~\cite{Zhou2015}. The importance of phasic GSR to trust sensing was also supported by Khawaji's study in which the average of peak GSR values was affected by interpersonal trust~\cite{khawaji_using_2015}. 
\section{Model Training and Validation} \label{sec:modeling}
The selected features discussed in Section~\ref{sec:FeatureSelection} were considered as input variables for each of the trust sensor models; the output variables were the categorical trust level, namely the classes `Trust' and `Distrust'. In this section we introduce the training procedure of a quadratic discriminant classifier that was used to predict the categorical trust class using the \psycho~features. We then present and discuss the results of the model validation. 

\subsection{Classifier Training}
The quadratic discriminant classifier was implemented using the Statistics and Machine Learning Toolbox in MATLAB R2016a (The MathWorks, Inc., USA). The low training and prediction time of quadratic discriminant classifiers is advantageous for real-time implementation of the classifier~\cite{mathworks_statistics_2016}. Moreover, the posterior probability calculated by the classifier for the class `Trust' was used as the probability of trust response, thus resulting in a continuous output. The continuous output of probability of trust response would be particularly beneficial for implementation of a feedback control algorithm for managing human trust level in an intelligent system. In order to avoid large and sudden fluctuations in the trust level, the continuous output was smoothed using a median filter with a window of size 15. The general trust sensor model and customized trust sensor models were developed with the same training procedure but with different feature sets (i.e., input variables). The former was based on the common feature set, and the latter was based on customized feature sets, as described in Sections~\ref{subsubsec:common-feature} and~\ref{subsubsec:custom-feature}. 

\subsection{Model Validation Techniques}
\begin{figure}
\centerline{\includegraphics[width=0.4\textwidth]{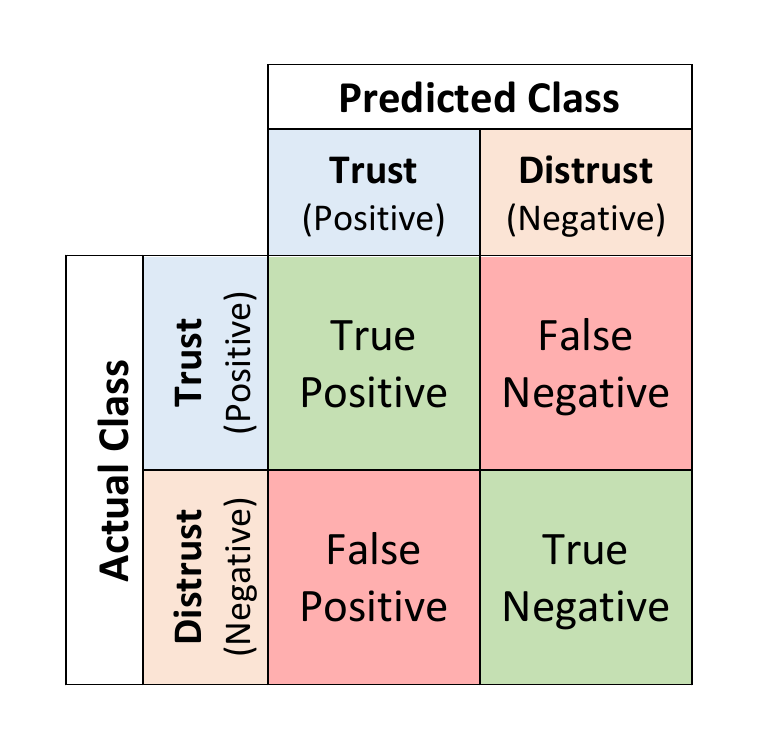}}
\caption{The actual class and the predicted class form a 2 $\times$ 2 confusion matrix. The outcomes are defined as true or false positive/negative.}
\label{fig:ConfusionMatrix}
\end{figure}

We used 5-fold cross-validation to evaluate the performance of classifiers.  The data, consisting of approximately 129 samples for each participant, was randomly divided into 5 sets. Each set was predicted using a model trained from the other four datasets. We used these predictions to evaluate the accuracy of the binary classification. Accuracy is defined as the proportion of correct predictions among the total number of samples and is given as

\begin{equation} \label{eq:accuracy}
\text{accuracy}=\frac{\text{Correct Predictions}}{\text{Total population}} \enspace .
\end{equation}
Moreover, prediction performance of a classifier may be better evaluated by examining the confusion matrix shown in Figure~\ref{fig:ConfusionMatrix}. We calculated two statistical measures called sensitivity
(true positive ratio) and specificity (true negative
ratio) that are defined as follows.

\begin{enumerate}

\item Sensitivity: the proportion of actual trust (positives) that are correctly predicted as such, where
\begin{equation} \label{eq:sensitivity}
\text{sensitivity}=\frac{\text{True positives}}{\text{True positives}+\text{False negatives}} \enspace .
\end{equation}

\item Specificity: the proportion of actual distrust (negatives) that were correctly predicted as such, where
\begin{equation} \label{eq:specificity}
\text{specificity}=\frac{\text{True negatives}}{\text{True negatives}+\text{False positives}} \enspace .
\end{equation}

\end{enumerate}

In order to examine the robustness of the classifier to the variation in training data, we performed 10,000 iterations with a different random division of the five sets in each iteration and calculated the performance measures for each iteration. Table~\ref{tab:result_common_insample} and Table~\ref{tab:result_common_outsample} show the mean, maximum (Max), minimum (Min), and standard deviation (SD) values for each of the performance measures for the \emph{general trust sensor model}. This is shown for both training-sample participants (Table IV) and validation-sample participants (Table V) along with the 95\% confidence interval (CI) obtained using the iterations.  Table~\ref{tab:result_custom} shows the performance statistics of the \emph{customized trust sensor model} for all participants. The confidence intervals obtained for both models were very narrow, \emph{indicating that models were robust to the selection of training data}.
\begin{table}
\caption{The accuracy, sensitivity, and specificity (\%) of the \emph{general} trust sensor model for training-sample participants with a 95\% confidence interval\label{tab:result_common_insample}}{
\begin{tabular}{lccc} 
 \hline
 		& Accuracy 			& Sensitivity 		& Specificity		\\ 
 \hline
Mean	& $70.52\pm0.007$	& $64.17\pm0.010$	& $75.49\pm0.009$	\\ 
Max		& $93.72\pm0.013$	& $96.75\pm0.020$	& $96.38\pm0.015$	\\ 
Min		& $54.67\pm0.042$	& $31.18\pm0.040$	& $44.92\pm0.039$	\\ 
SD		& $11.29\pm0.006$	& $18.96\pm0.009$	& $14.35\pm0.008$	\\ 
 \hline
\end{tabular}}
\end{table}

\begin{table}
\caption{The accuracy, sensitivity, and specificity (\%) of the \emph{general} trust sensor model for validation-sample participants with a 95\% confidence interval\label{tab:result_common_outsample}}{
\begin{tabular}{lccc} 
 \hline
 		& Accuracy 			& Sensitivity 		& Specificity		\\ 
 \hline
Mean	& $73.13\pm0.010$	& $65.35\pm0.015$	& $79.49\pm0.013$	\\ 
Max		& $99.89\pm0.006$	& $99.92\pm0.006$	& $99.85\pm0.011$	\\
Min		& $59.29\pm0.035$	& $34.35\pm0.081$	& $57.04\pm0.050$	\\
SD		& $10.91\pm0.007$	& $17.03\pm0.016$	& $12.26\pm0.015$	\\
 \hline
\end{tabular}}
\end{table}

\begin{table}
\caption{The accuracy, sensitivity, and specificity (\%) of the \emph{customized} trust sensor model for all participants with a 95\% confidence interval\label{tab:result_custom}}{
\begin{tabular}{lccc} 
 \hline
 		& Accuracy 			& Sensitivity 		& Specificity		\\
 \hline
Mean	& $78.55\pm0.005$	& $72.83\pm0.007$	& $82.56\pm0.007$	\\
Max		& $100.00\pm0.000$	& $100.00\pm0.000$	& $100.00\pm0.000$	\\
Min		& $61.59\pm0.041$	& $34.77\pm0.044$	& $45.89\pm0.040$	\\
SD		& $9.69\pm0.005$	& $17.02\pm0.008$	& $11.18\pm0.007$	\\
 \hline
\end{tabular}}
\end{table}

\subsection{Discussion on Performance of Classification Models}
The mean accuracy was 70.52$\pm$0.007\% for training-sample participants.  Similarly, the mean accuracy for the \emph{validation-sample} participants was 73.13$\pm$0.010\%. The fact that the performance of the general trust model was consistent for both training-sample and validation-sample participants suggests that the identified list of features could estimate trust for a broad population of individuals. Moreover, the mean accuracy was 78.58$\pm$0.0005\% for the customized trust sensor models for all participants. Recall that the customized trust senor models were based on a customized feature set for each participant. There were 12 significant features to predict trust for the general trust sensor models, while less than 5 features were needed for the customized trust sensor models. These findings support the hypothesis that a customized trust sensor model could enhance the prediction accuracy with a smaller feature set.
For some individual participants, the mean accuracy increased to 100\%.

\begin{figure}
\centering
\subfigure[General Trust Sensor model predictions with an accuracy of 90.52\%.\label{subfig:general_good_grp1}]{\includegraphics[width=0.8\textwidth]{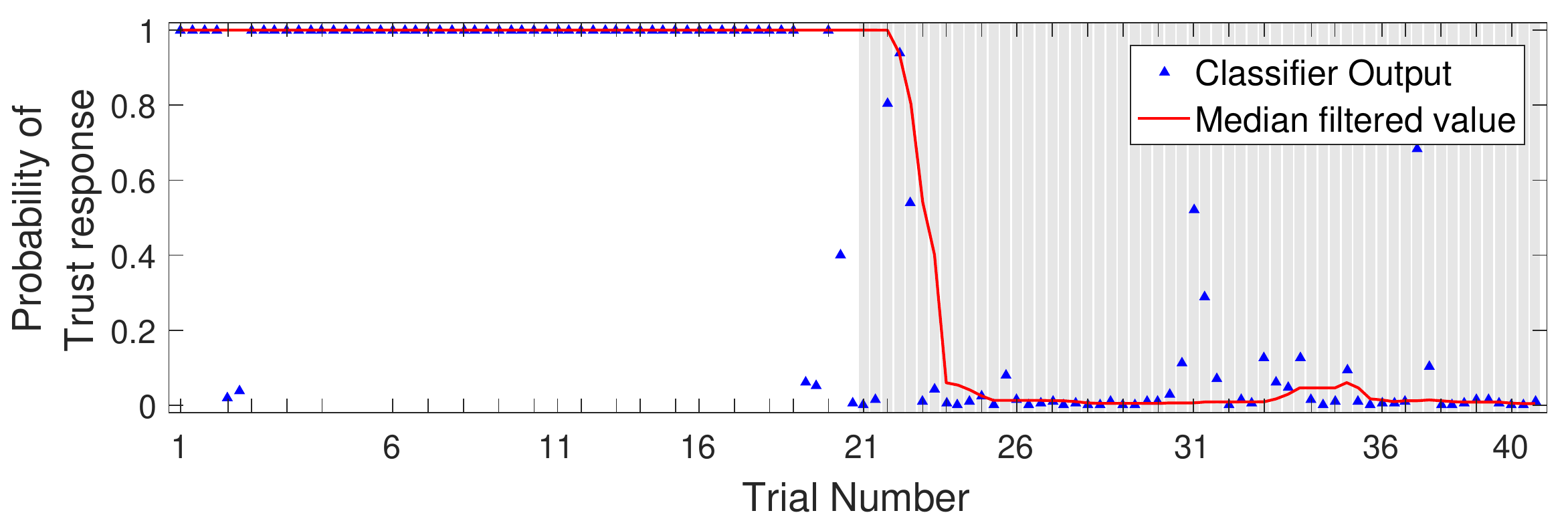}}

\subfigure[Customized Trust Sensor model predictions with an accuracy of 93.97\%.\label{subfig:custom_good_grp1}]
{\includegraphics[width=0.8\textwidth]{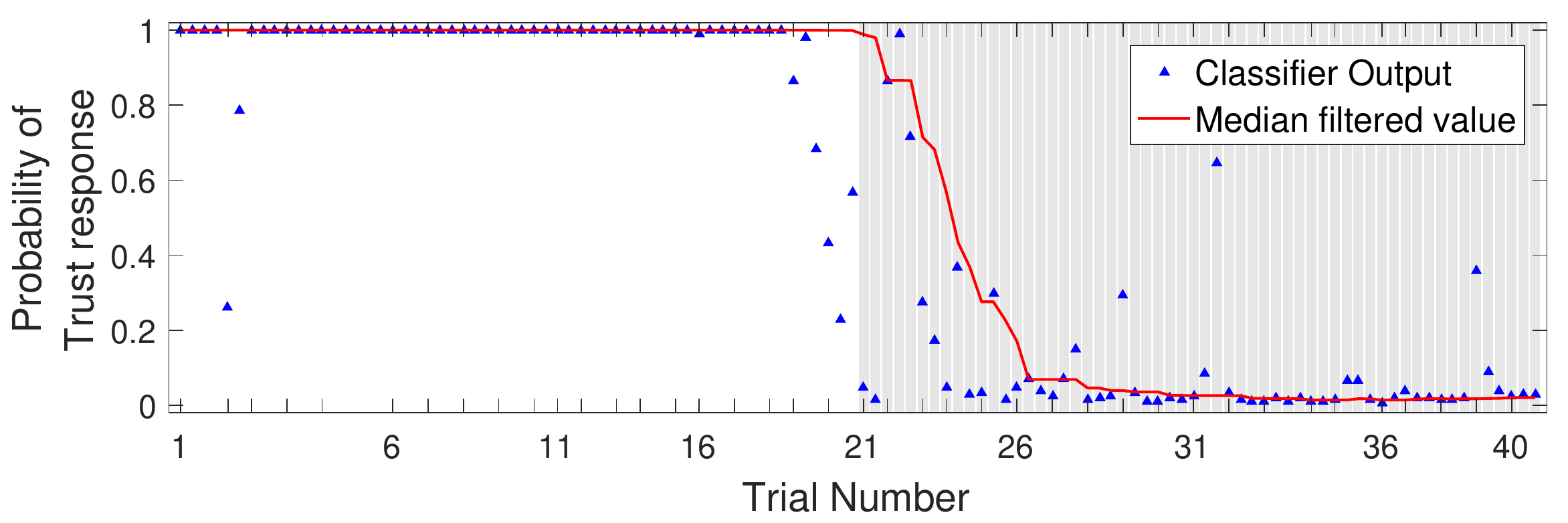}}
\caption{Classifier predictions for participant 44 in group 1. Faulty trials are highlighted in gray. Trust sensor models had a good accuracy for this participant. The classifier output of posterior probability was smoothed using a median filter with window of size 15.}
\label{fig:good_grp1}
\end{figure}

\begin{figure}
\centering
\subfigure[General Trust Sensor model predictions with an accuracy of 91.12\%.\label{subfig:general_good_grp2}]{\includegraphics[width=0.8\textwidth]{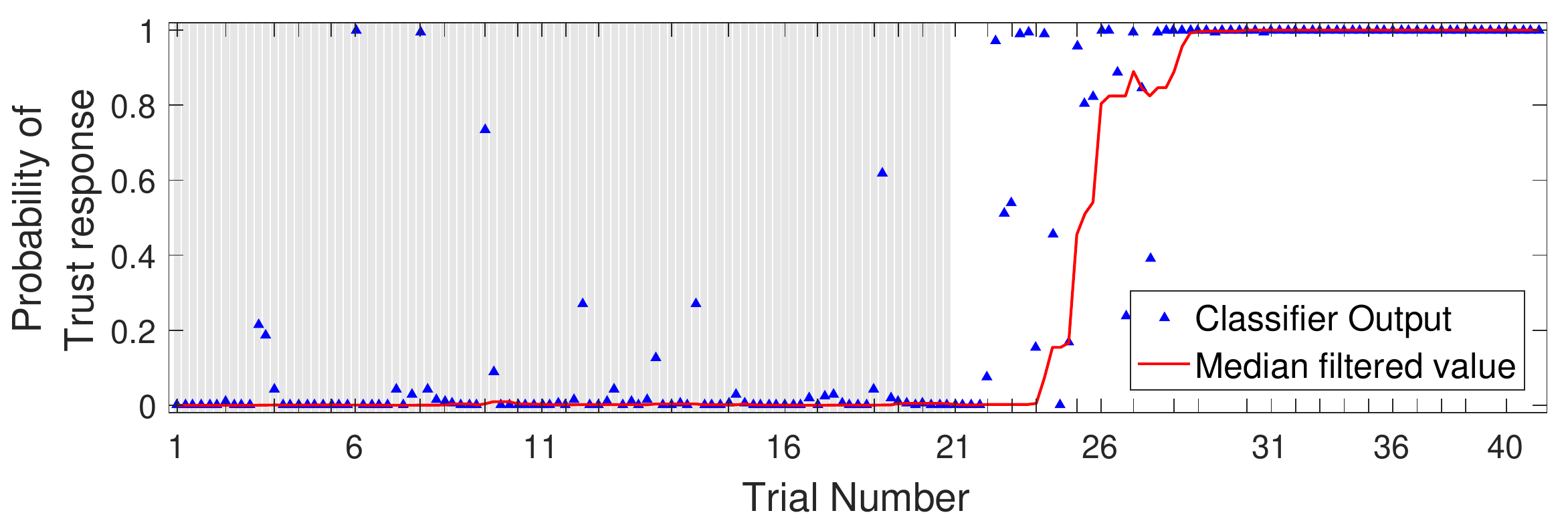}}

\subfigure[Customized Trust Sensor model predictions with an accuracy of 96.45\%.\label{subfig:custom_good_grp2}]
{\includegraphics[width=0.8\textwidth]{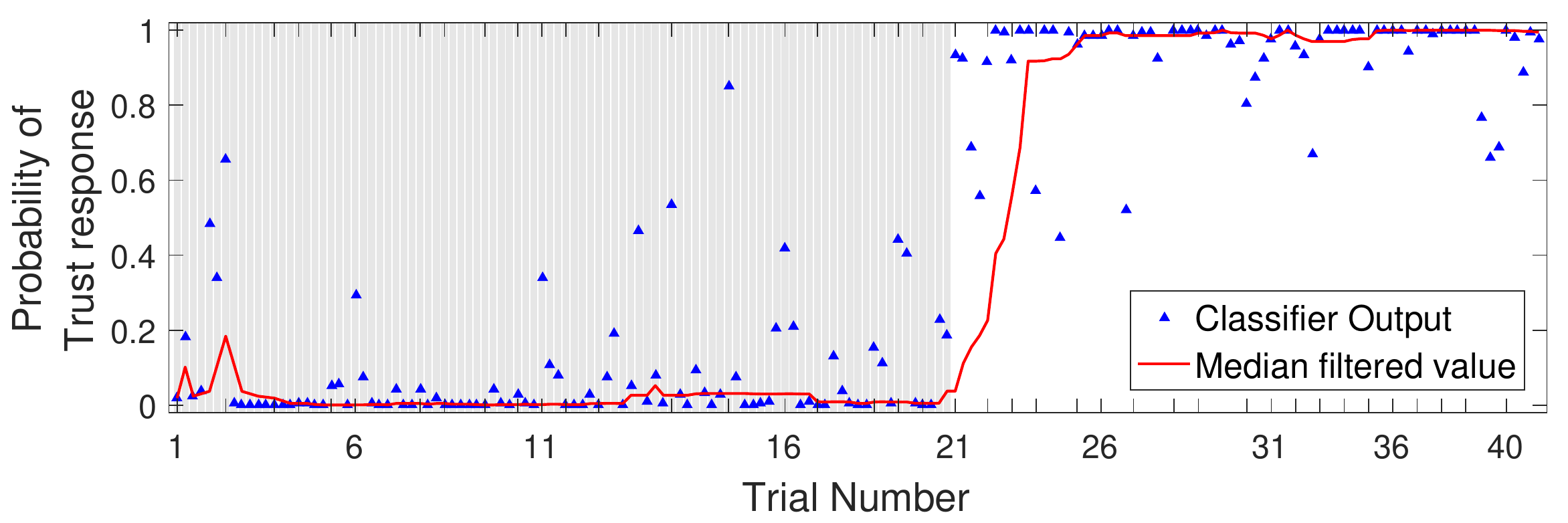}}
\caption{Classifier predictions for participant 10 in group 2. Faulty trials are highlighted in gray. Trust sensor models had good accuracy for this participant.  The classifier output of posterior probability was smoothed using a median filter with window of size 15.}
\label{fig:good_grp2}
\end{figure}

\begin{figure}
\centering
\subfigure[General Trust Sensor model predictions with an accuracy of 61.26\%.\label{subfig:general_bad_grp1}]{\includegraphics[width=0.8\textwidth]{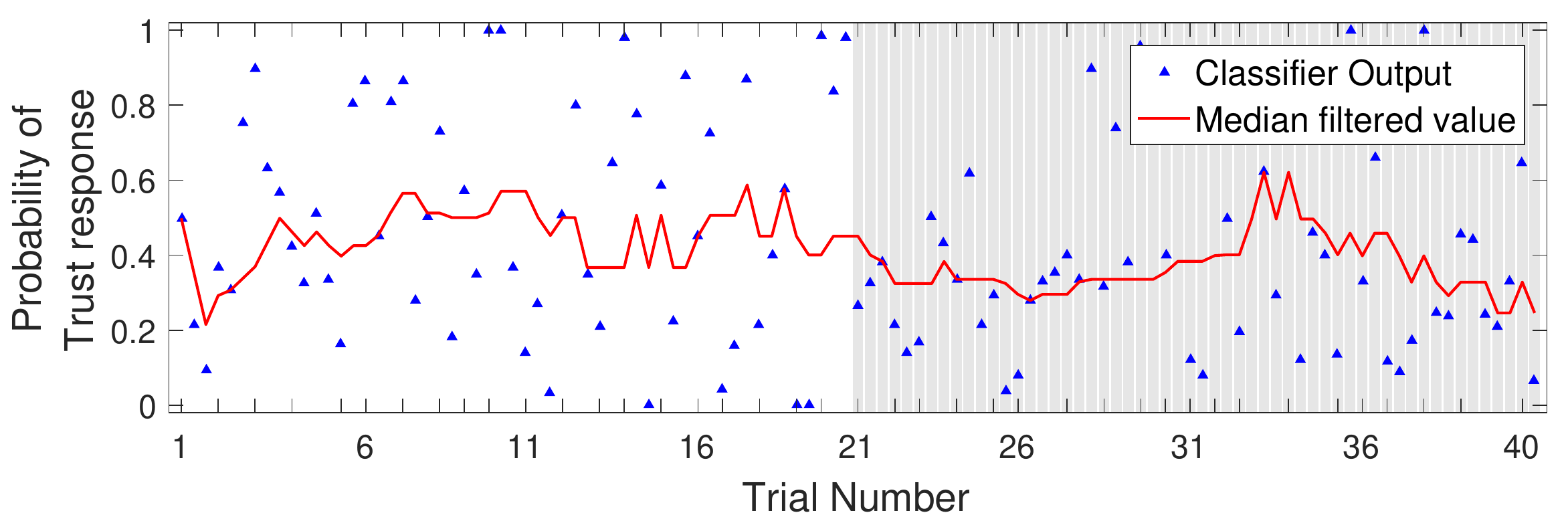}}

\subfigure[Customized Trust Sensor model predictions with an accuracy of 72.07\%.\label{subfig:custom_bad_grp1}]
{\includegraphics[width=0.8\textwidth]{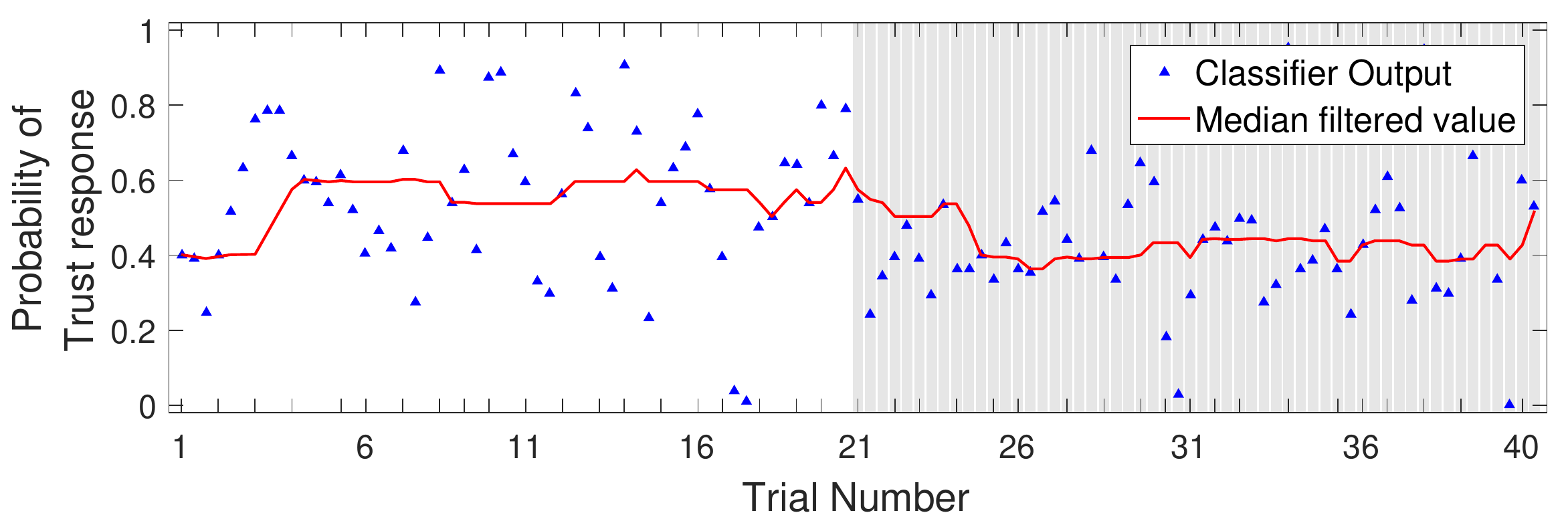}}
\caption{Classifier predictions for participant 8 in group 1. Faulty trials are highlighted in gray. Trust sensor models did not have good accuracy for this participant.  The classifier output of posterior probability was smoothed using a median filter with window of size 15.}
\label{fig:bad_grp1}
\end{figure}

Figures~\ref{fig:good_grp1} and~\ref{fig:good_grp2} are examples of good predictions for participants in groups 1 and 2, respectively. The customized trust sensor models performed better for both participants, specifically at the transition state at the beginning of database 2. Figure~\ref{subfig:custom_good_grp1} shows an example of a transition state at the beginning of database 2; it took five trials for this participant to establish a new trust level. The classification accuracy was low for some participants as shown in Figure~\ref{fig:bad_grp1}. The classifier had difficulty correctly predicting trust (database 1), which may imply that this particular participant was not able to conclude whether or not to trust the sensor report, even in reliable trials. Another potential reason could be that trust variations of this participant did not result in significant changes in their physiological signals. Nevertheless, the customized trust sensor model still showed a higher accuracy than the general trust sensor model. 

The general trust sensor model resulted in mean \emph{specificity} of 75.49$\pm$0.009\% and 79.49$\pm$0.013\% for training-sample and validation-sample participants, respectively. The customized trust sensor model resulted in 82.56$\pm$0.007\% for all participants. This indicates that the models are capable of correctly predicting distrust in humans. The models are less likely to predict a distrust response as trust (i.e., less false positives). The mean \emph{sensitivity} was 64.17$\pm$0.010\% and 65.35$\pm$0.015\% for the general trust sensor model for training-sample and validation-sample participants, respectively. The customized trust sensor model resulted in 72.83$\pm$0.007\% for all participants. Low sensitivity (more false negatives) occurs when the model often predicts trust as distrust. In the context of using this trust sensor model to design an intelligent system that could be responsive to a human's trust level, low sensitivity would arguably not have an adverse effect since the goal of the system would be to enhance trust. 

There is a fundamental trade-off that exists between the general and customized models in terms of the time spent on model training and model performance as shown in Table~\ref{tab:comparison}. The results show that the selected feature set (Table~\ref{tab:feature-selection}) for the general trust sensor models is applicable for a general adult population with a 71.22\% mean accuracy (i.e., the mean accuracy calculated across all participants).  Furthermore, by applying this common feature set, feature selection is not required while implementing the general model. This would reduce the model training time and potentially make the model adaptable to various scenarios. However, the common feature set for a general population is larger than feature sets optimized for each individual because it attempts to accommodate an aggregated group of individuals. Therefore, in scenarios where the speed of the online prediction process is the priority, the customized trust sensor model, with a smaller feature set, would be preferred. The customized trust sensor model also enhances the prediction accuracy. Nonetheless, it is worth noting that implementing the customized trust sensor model would still require extraction of a larger set of features initially for training followed by a smaller feature set extraction for real-time implementation. This would increase the time required for training the model as an additional feature selection step would need to be performed. 

While we focused on situational and learned trust, dispositional trust factors, such as demographics, may have partially contributed to the observed lower accuracy of the general trust sensor model due to individual differences in trust response behavior~\cite{riedl2010,Akash2017}. Incorporating these additional factors and other \psycho~signals may increase the trust estimation accuracy of the trust sensor model, as the features included in the present model inherently represent only a subset of many non-verbal signals that correlate to trust level.

In summary, the proposed trust sensor model could be used to enable intelligent systems to estimate human trust and in turn respond to, and collaborate with, humans in such a way that leads to successful and synergistic collaborations. Potential human-machine/robot collaboration contexts include robotic nurses that assist patients, aircrafts that exchange control authority with human operators, and numerous others~\cite{wang2017trends}.

\begin{table}
\caption{Comparison of General Trust Sensor Model and Customized Trust Sensor Model for implementation\label{tab:comparison}}{
\begin{tabular}{lcc} 
 \hline
 Model Characteristics		& General Trust Sensor Model	& Customized Trust Sensor Model \\ 
 \hline
Required training time 		& Less 							& More 							\\
Size of final feature set 	& 12 							& 	4.33 (Average) 				\\
Prediction Time 			& More	 						& Less 							\\
Mean Prediction Accuracy 	& 71.22\%  						& 78.55\% 						\\
 \hline
\end{tabular}}
\end{table}
\section{Conclusion} \label{sec:conclusion}
As humans are increasingly required to interact with intelligent systems, trust becomes an important factor for synergistic interactions. The results presented in this paper show that \psycho~measurements can be used to estimate human trust in intelligent systems in real-time. By doing so, intelligent systems will have the ability to respond to changes in human trust behavior.

We proposed two approaches for developing classifier-based empirical trust sensor models that estimate human trust level using \psycho~measurements. These models used human subject data collected from 45 participants. The first approach was to consider a common set of \psycho~features as the input variables for any human and train a classifier-based model using this feature set, resulting in a general trust sensor model with a mean accuracy of 71.22\%. The second approach was to consider a customized feature set for each individual and train a classifier-based model using that feature set; this resulted in a mean accuracy of 78.55\%. The primary trade-off between these two approaches was shown to be training time and performance (based on mean accuracy) of the classifier-based model.  That is to say, while it is expected that using a feature set customized to a particular individual will outperform a model based upon the general feature set, the time needed for training such a model may be prohibitive in certain applications. Moreover, although the criteria used for feature selection and classifier training in this study was mean accuracy, a different criterion could be chosen to adapt to various applications. Finally, future work will involve increasing the sample size and augmenting the general trust sensor model to account for dispositional trust factors in order to improve the prediction accuracy of the model. It will also be important to test the established framework in both simulated and immersive environments using, for example, driving or flight simulators and/or virtual reality, as well as in real-life settings.

\bibliographystyle{ACM-Reference-Format}
\bibliography{Reference}

\end{document}